\documentclass[sigconf]{acmart}

\newif\ifextendedversion
\extendedversiontrue % \extendedversionfalse for workshop version (6 pages), \extendedversiontrue for arxiv version

\copyrightyear{2024}
\acmYear{2024}
\setcopyright{cc}
\acmConference[DIN '24]{Proceedings of the ACM Conext-2024 Workshop on the Decentralization of the Internet}{December 9--12, 2024}{Los Angeles, CA, USA}
\acmBooktitle{Proceedings of the ACM Conext-2024 Workshop on the Decentralization of the Internet (DIN '24), December 9--12, 2024, Los Angeles, CA, USA}
\acmDOI{10.1145/3694809.3700740}
\acmISBN{979-8-4007-1252-4/24/12}

\begin{document}
\title{Bluesky and the AT Protocol: Usable Decentralized Social Media}
\ifextendedversion
\titlenote{This is an extended version of a paper appearing at the ACM Conext-2024 Workshop on the Decentralization of the Internet (DIN '24). Section~\ref{sec:moderation} and Section~\ref{sec:related-work} of this version are omitted in the workshop version.}
\settopmatter{printfolios=true,printccs=false} % extended version has page numbers
\fi

\author{Martin Kleppmann}
\email{martin.kleppmann@cst.cam.ac.uk}
\orcid{0000-0001-7252-6958}
\affiliation{\institution{University of Cambridge}\city{Cambridge}\country{UK}}

\author{Paul Frazee}
\email{paul@blueskyweb.xyz}
\orcid{0009-0005-6762-4357}
\affiliation{\institution{Bluesky Social PBC}\city{Seattle}\state{WA}\country{USA}}

\author{Jake Gold}
\email{jake@blueskyweb.xyz}
\orcid{0009-0009-1717-4967}
\affiliation{\institution{Bluesky Social PBC}\city{Mountain View}\state{CA}\country{USA}}

\author{Jay Graber}
\email{jay@blueskyweb.xyz}
\orcid{0009-0003-1273-3308}
\affiliation{\institution{Bluesky Social PBC}\city{Seattle}\state{WA}\country{USA}}

\author{Daniel Holmgren}
\email{daniel@blueskyweb.xyz}
\orcid{0009-0007-4538-1707}
\affiliation{\institution{Bluesky Social PBC}\city{Springfield}\state{MO}\country{USA}}

\author{Devin Ivy}
\email{devin@blueskyweb.xyz}
\orcid{0009-0008-4240-8676}
\affiliation{\institution{Bluesky Social PBC}\city{Cambridge}\state{MA}\country{USA}}

\author{Jeromy Johnson}
\email{why@blueskyweb.xyz}
\orcid{0009-0004-3161-9890}
\affiliation{\institution{Bluesky Social PBC}\city{Seattle}\state{WA}\country{USA}}

\author{Bryan Newbold}
\email{bryan@blueskyweb.xyz}
\orcid{0000-0002-8593-9468}
\affiliation{\institution{Bluesky Social PBC}\city{Seattle}\state{WA}\country{USA}}

\author{Jaz Volpert}
\email{jaz@blueskyweb.xyz}
\orcid{0009-0009-2935-6020}
\affiliation{\institution{Bluesky Social PBC}\city{San Francisco}\state{CA}\country{USA}}

\begin{abstract}
    Bluesky is a new social network built upon the AT Protocol, a decentralized foundation for public social media.
    It was launched in private beta in February 2023, and has grown to over 10 million registered users by October 2024.
    In this paper we introduce the architecture of Bluesky and the AT Protocol, and explain how the technical design of Bluesky is informed by our goals: to enable decentralization by having multiple interoperable providers for every part of the system; to make it easy for users to switch providers; to give users agency over the content they see; and to provide a simple user experience that does not burden users with complexity arising from the system's decentralized nature.
    The system's openness allows anybody to contribute to content moderation and community management, and we invite the research community to use Bluesky as a dataset and testing ground for new approaches in social media moderation.
\end{abstract}



\maketitle
\renewcommand{\shortauthors}{Martin Kleppmann et al.}

\ifextendedversion\newpage\fi

\section{Introduction}\label{sec:introduction}

Over the last two decades, social media has evolved into a ``digital town square'' and a cornerstone of civic life~\cite{Barabas:2017}.
When a social media service is under the full control of a single corporation, it may change its policies on the whim of its leaders~\cite{Yeung:2023}.
Its operations are opaque (e.g.\ regarding which content is recommended to users), and its users lack agency over their user experience.
As a result, there has been increasing interest in decentralized social networks, of which the \emph{fediverse} around the ActivityPub protocol~\cite{ActivityPub} and the Mastodon software~\cite{Mastodon} is perhaps the best known.

However, decentralization also introduces new challenges.
For example, in the case of Mastodon, a user needs to choose a server when creating an account.
This choice is significant because the server name becomes part of the username; migrating to another server implies changing username, and preserving one's followers during such a migration requires the cooperation of the old server.
If a server is shut down without warning, accounts on that server cannot be recovered~-- a particular risk with volunteer-run servers.
In principle, a user can host their own server, but only a small fraction of social media users have both the technical skills and the inclination to do so.

The distinction between servers in Mastodon introduces complexity for users that does not exist in centralized services.
For example, a user viewing a thread of replies in the web interface of one server may see a different set of replies compared to viewing the same thread on another server, because a server only shows those replies that it knows about~\cite{Adida:2022}.
As another example, when viewing the web profile of an account on another server, clicking the ``follow'' button does not simply follow that account; instead, the user needs to enter the hostname of their own server and be redirected to a URL on their home server before they can follow the account.
In our opinion, it is undesirable to burden users with such complexity arising from the federated architecture.

In this paper we introduce the \emph{AT Protocol} (atproto), a decentralized foundation for social networking, and \emph{Bluesky}, a Twitter-style microblogging app built upon it.
A core design goal of atproto and Bluesky is to enable a user experience of the same or better quality as centralized services, while being open and decentralized on a technical level.
%We introduce the user-facing features of Bluesky in Section~\ref{sec:product}, and in Section~\ref{sec:architecture} we explain the underlying systems architecture.
The AT Protocol is designed such that for every part of the system there can be multiple competing operators providing interoperable services, making it easy to switch providers.

Decentralization alone is not able to solve some of the thorniest problems of social media, such as misinformation, harassment, and hate speech~\cite{Roth:2023}.
However, by opening up the internals of a service to the public, decentralization can enable a marketplace of approaches to these problems~\cite{Masnick:2019}.
For example, Bluesky allows anybody to run moderation services that make subjective decisions of selecting desirable content or flagging undesirable content, and users can choose which moderation services they want to subscribe to.
Moderation services are decoupled from hosting providers, making it easy for users to switch moderation services until they find ones that match their preferences.
We hope that this architectural openness enables communities to develop their own approaches to managing problematic content, independently of what hosting providers implement~\cite{Masnick:2019}.

For example, researchers wanting to identify disinformation campaigns can easily get access to all content being posted, the social graph, and user profiles on Bluesky.
If they are able construct an algorithm to label suspected disinformation, they can publish their labels in real time, and users who wish to see those labels can enable them in their client.
One goal of this paper is to bring Bluesky and the AT Protocol to the attention of researchers working on such algorithms, and to invite them to use the rapidly growing dataset of Bluesky content as a basis for their work.

%\begin{figure}
%   \centering
%   \includegraphics[width=0.7\linewidth]{home-feed.png}
%   \Description{A Twitter-like feed of short posts. At the top is a feed selector, in which the default ``Following'' feed is active.}
%   \caption{Screenshot of the Bluesky home screen.}
%   \label{fig:home-feed}
%\end{figure}

\section{The Bluesky Social App}\label{sec:product}

Bluesky is a microblogging application in the style of Twitter/X. % (see Figure~\ref{fig:home-feed}).
The ``official'' client app is available on iOS, Android, and the web; several independently developed client apps are also available, such as Graysky~\cite{Graysky} and deck.blue~\cite{deck.blue}.
Users can make public posts containing up to 300 characters of text, up to four images, and short-form video.
They can interact with posts by replying, reposting, or liking, and they can follow other users.
By default users have two feeds: one showing posts by followed accounts in reverse chronological order, and one recommending popular content across the network.
Users can also choose from many alternative feeds that show content on various topics, as explained below.

%\begin{figure}
%   \centering
%   \includegraphics[width=\linewidth]{user-growth.pdf}
%   \Description{An exponential growth curve, approximately doubling every month, starting at 55k in May 2023 and exceeding 1.3M in October 2023.}
%   \caption{Number of registered users on Bluesky since April 2023.}
%   \label{fig:user-growth}
%\end{figure}

Bluesky launched an invite-only beta in February 2023, and has grown to over 10~million registered users in the following 20 months. % as shown in Figure~\ref{fig:user-growth}.
Bluesky Social PBC (a public-benefit corporation) develops the official client app and operates the core services; the client and several server-side components are open source, dual-licensed under the MIT and Apache 2.0 licenses~\cite{BlueskyGithub}.
The protocols they use are defined by open specifications~\cite{AtProtoSpecs}.
Several parts of the system, such as feed generators and alternative clients~\cite{AtProtoClients} are developed and operated by independent third parties.

\ifextendedversion
\subsection{Moderation Features}\label{sec:moderation}

Bluesky currently has the following moderation mechanisms (additional mechanisms are under discussion~\cite{Moderation}):
\begin{description}
    \item[Content filtering:] Automated systems label potentially problematic content (such as spam, or images of a sexual or violent nature), and the app's preferences allow users to choose whether to show or hide content in each of these categories in their feeds. Some filters are always applied regardless of user preferences.
    \item[Mute:] A user can mute specific accounts or threads, which hides the muted content from their own feeds and notifications. The content continues to be visible to other users, and the target does not know that they were muted. A user can also publish a mutelist of accounts, and other users can subscribe to that list, which has the same effect as if they individually muted all of the accounts on the list.
    \item[Block:] One user can block another, which prevents all future interactions (such as mentions, replies, or reposts) between those accounts in addition to bidirectional muting. Similarly to mutelists, a user can also publish a list of accounts, and other users can block all accounts on that list by subscribing to it.
    \item[Interaction gating:] A user who makes a post can restrict who is allowed to reply to it (anyone, anyone they follow, anyone mentioned in the post, and/or anyone on a particular list of accounts)~\cite{ReplyGating}.
        The original poster can also hide replies to be accessible but not shown by default.
    \item[Quote detachment:] When a post is quoted by another user's post, the author of the original post can detach it from the quote post.
        This reduces the risk of harassment through dog-piling.
    \item[Takedown:] Users can report content that violates the terms of service to moderation services.
        Operators of the services described in Section~\ref{sec:architecture} can take down violating media, posts, or accounts.
    \item[Custom feeds:] While the aforementioned mechanisms provide negative moderation (helping users avoid content they do not want to see), feed generators (see Section~\ref{sec:feeds}) can actively select high-quality content.
\end{description}
\fi % \ifextendedversion

% full account takedown at PDS: easy, entire repo unavailable
% take down a post at PDS: one record is unavailable -- is not announced in event stream, not removed from MST, not filtered from full repo download. per-post takedown at app view makes more sense
% take down a blob at PDS: reference stays there, but refuse to serve the blob (quarantine it, move out of user's storage)

\ifextendedversion\subsection{User Handles}\else\paragraph{User Handles}\fi

Like on Twitter/X, a Bluesky user has two names: the \emph{display name} can be almost any string, and the \emph{handle} needs to uniquely identify a user.
A handle, prefixed with an @ sign, is used to mention another user in a post.
% Examples can be seen in Figure~\ref{fig:home-feed} (the display name is in bold, and the handle is in a smaller font and lighter color).

The need for handles to be unique creates challenges in decentralized systems, since it requires an authority that determines which handle is assigned to which user.
Mastodon's approach is to include the server name in the handle, which makes it difficult to move to another server.
An alternative would be to use a blockchain-based naming system, such as the Ethereum Name System (ENS)~\cite{ENS}; this has the disadvantage of requiring the user to buy cryptocurrency in order to create an account, which we wanted to avoid.

Instead, Bluesky and atproto use DNS domain names as handles.
Users can sign up for a subdomain of \texttt{.bsky.social} for free.
If a user already owns a domain name, they can claim it as their Bluesky handle by adding a DNS record or by hosting a file under a \texttt{/.well-known/} HTTPS URL on that domain~\cite{DomainHandle}.
Users can also buy a new domain name within Bluesky, via a partnership with a domain registrar~\cite{PurchaseDomain}.
Domain names have several advantages:
\begin{itemize}
    \item We leverage the existing infrastructure of ICANN, registrars, and name servers, including for example the dispute resolution procedures for trademarks.
    \item Domain names are a well-known concept even among non-technical users, and they are short and simple.
    \item A user can move to a different server without changing their handle (see Section~\ref{sec:identity}).
    \item Users do not need to host their own server to use their own domain name; they only need to maintain a DNS record.
    \item For organizations and people that already have a well-known domain name, using that name makes it easy for users to check that their Bluesky account is genuine. For example, the New York Times' handle is \texttt{@nytimes.com}.
    \item An organization can allow its staff to demonstrate their affiliation by granting them subdomains of the organization's main domain name (comparable to institutional email addresses). For example, a journalist's handle may indicate that they are at a particular news organization.
    \item Offering free subdomains costs very little.
\end{itemize}

\ifextendedversion\subsection{Custom Feeds and Algorithmic Choice}\label{sec:feeds}\else\paragraph{Custom Feeds and Algorithmic Choice}\fi

Several decentralized social networks offer only a reverse-chrono\-logical feed of posts from accounts the user is following~-- a backlash against the opaque recommendation algorithms used by mainstream social networks.
For example, Mastodon advertises itself as having ``no algorithms or ads to waste your time''~\cite{Mastodon}.

Our belief is that the problem lies not with algorithms \emph{per se}, but rather with centrally controlled, opaque algorithms that remove user agency and prioritize user engagement over all else, e.g.\ by promoting controversial posts.
Good recommendation algorithms can help users discover content that is relevant to them and find new accounts to follow~-- especially important for new users who are not yet following many accounts.
They are also helpful for surfacing content on a particular topic, whereas following a user means seeing all of their posts, which might be on a mixture of topics, not all necessarily interesting to all followers.
Our goal is to offer an open and diverse marketplace of algorithms in which communities can adapt the system to suit their needs, and users have more agency over how they spend their time and attention~\cite{AlgorithmicChoice}.
% Giving users the ability to choose their algorithms lets them control what they want to see, rather than having the platform decide for them.

Bluesky Social PBC offers a selection of feed algorithms of its own, and also allows anybody to create their own feed generator~\cite{CustomFeeds}.
Tens of thousands of custom feeds have already been created.
% Section~\ref{sec:labeling} explains how feed generators work.
% In Figure~\ref{fig:home-feed}, a selection of bookmarked feeds is given at the top of the screen; in this example, the selected ``Following'' feed is the default reverse-chronological timeline, while ``Week Peak Feed'' (network-wide posts with many likes from the last week) and ``Birds!'' (photos and posts from birdwatchers) are third-party feeds.
A feed generator can use arbitrary criteria to select its content.
For example, it may use a manually curated list of accounts, and select posts from those accounts that contain a particular hashtag or emoji character.
Use of machine learning algorithms is equally possible.

\section{The AT Protocol Architecture}\label{sec:architecture}

Bluesky is the social app with the features explained in Section~\ref{sec:product}, while the AT Protocol is the underlying decentralized foundation.
We maintain this separation because the AT Protocol is designed to support multiple \emph{social modes}, not just Bluesky.
For example, besides a Twitter-style microblogging app, atproto could also be used to implement Reddit-style forums, long-form blogs with comments, or domain-specific social applications such as link sharing or book reviews.
The same user identity, social graph, and user data storage servers can be shared between all of these apps.
% https://whtwnd.com/ is a blogging platform on top of atproto
% open sourcing announcement: https://whtwnd.com/whtwnd.com/3kutsnrkgvk2o
% https://frontpage.fyi/ is a HN clone
% https://turntabl.app/
% https://smokesignal.events/ is an events management system
% Ask Jaz for more

A social mode is defined by a \emph{lexicon}, which specifies the schema of the data and the request endpoints involved~\cite{AtProtoSpecs}.
Several lexicons are currently used: for example, the \texttt{com.atproto} lexicon defines the core AT Protocol concepts such as user identity (Section~\ref{sec:identity}), the \texttt{app.bsky} lexicon defines the microblogging mode, and the \texttt{tools.ozone} lexicon supports moderation actions~\cite{BlueskyLexicons}.
Anyone can define a new lexicon, allowing new social modes to coexist alongside the Bluesky social app on a shared infrastructure.
The purpose of a lexicon is to provide documentation, to allow code generation and type-checking in applications, and to facilitate schema evolution for long-term compatibility and interoperability.

The biggest constraint for new social modes is that atproto is currently designed for content that users want to make public.
In particular, Bluesky user profiles, posts, follows, and likes are all public.
Blocking actions are also currently public; we are investigating mechanisms for making these private~\cite{PublicBlocks,PrivateBlocks}.
Only a small amount of user state is currently private: muted accounts and threads, notifications and their unread status, and user preferences such as content filtering settings.
Private communications (direct messages) are currently handled by a centralized service run by Bluesky Social PBC; we plan to decentralize this component in the future.

\begin{figure*}
    \centering
    \includegraphics[width=\linewidth]{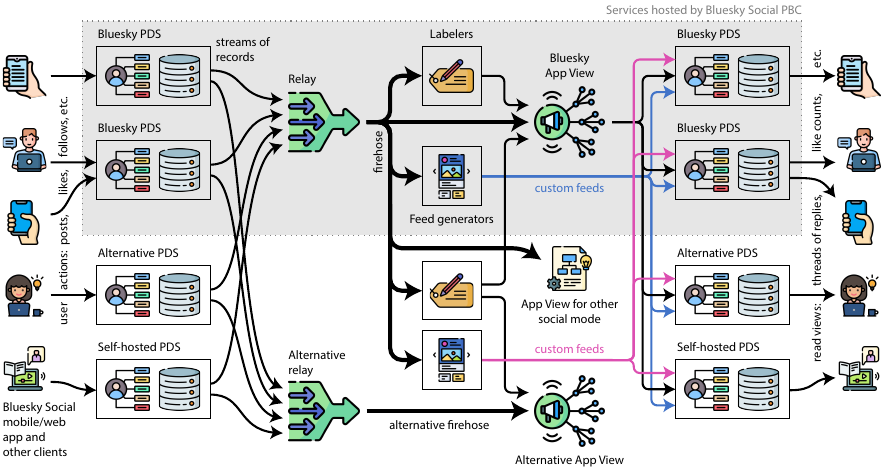}
    \Description{Client apps send posts, likes, etc. to the user's home PDS. A stream of records from each PDS is consumed by the Relay, and aggregated to form the firehose. The firehose is consumed by the Bluesky App View as well as labelers and feed generators. PDSes read from the App View to obtain threads of replies, like counts, etc. Bluesky Social PBC currently hosts several PDSes, the Relay, the App View, and some of the labelers and feed generators. The Bluesky App View also pulls data from external labelers and feed generators. Other providers can set up alternative PDSes, aggregators, and app views if they wish (including app views for social modes other than microblogging).}
    \caption{The main services involved in providing Bluesky, and data flows between them. Icons from Flaticon.com.}
    \label{fig:indexing}
\end{figure*}

% Icons used:
% https://www.flaticon.com/free-icon/browsing_2975669
% https://www.flaticon.com/free-icon/mobile_6300115
% https://www.flaticon.com/free-icon/segmentation_6012134
% https://www.flaticon.com/free-icon/server_689360
% https://www.flaticon.com/free-icon/simplify_5271304
% https://www.flaticon.com/free-icon/product-development_4229803
% https://www.flaticon.com/free-icon/carousel_9893850
% https://www.flaticon.com/free-icon/virality_11940519
% https://www.flaticon.com/free-icon/content-plan_11940521
% https://www.flaticon.com/free-icon/blogger_3893160
% https://www.flaticon.com/free-icon/online-learning_12641520
% https://www.flaticon.com/free-icon/working_5654681

\subsection{User Data Repositories}\label{sec:repos}

All data that a user wishes to publish is added to their \emph{repository}, which stores a collection of \emph{records}.
Whenever a user performs some action~-- making a post, liking another user's post, following another user, etc.~-- that action becomes a record in their repository.
Records are encoded in DAG-CBOR~\cite{DAG-CBOR}, a restricted form of CBOR~\cite{CBOR}, a compact binary data format.
The schema of records is defined by the lexicon, and a repository may contain a mixture of records from several different lexicons, representing user actions in different social modes.
Media files (e.g.\ images) are stored outside of the user's repository, but referenced by their CID~\cite{CID} (essentially a cryptographic hash) from a record in the repository.
Similarly, a reference to a record in another repository (e.g.\ identifying a post being liked) also includes its CID.

Each user account has one repository, and it contains all of the actions they have ever performed, minus any records they have explicitly deleted.
A \emph{Personal Data Server (PDS)} hosts the user's repository and makes it publicly available as a web service; we discuss PDSes in more detail in Section~\ref{sec:pds}.

A user only updates their own repository; for example, if user $A$ follows user $B$, this results only in a follow record in user $A$'s repository, and no change to $B$'s repository.
To find all followers of user $B$ requires indexing the content of all repositories.
This design decision is similar to the way hyperlinks work on the web: it is easy to find all the outbound links from a web page at a given URL, but to find all the inbound links to a page requires an index of the entire web, which is maintained by web search engines.

The \emph{AT} in atproto stands for \emph{Authenticated Transfer}, which reflects the fact that repositories are cryptographically authenticated.
The records in a repository are organized into a \emph{Merkle Search Tree} (MST), a type of Merkle tree that remains balanced, even as records are inserted or deleted in arbitrary order~\cite{Auvolat:2019}.
After every change to a repository, the root hash of the MST is signed; the public verification key for this signature is part of the user identity described in Section~\ref{sec:identity}.
This enables an efficient cryptographic proof that a given record appears within a given user's repository.
Moreover, when a user updates or deletes a record, the MST enables a proof that the old record no longer appears in the repository.

\subsection{Personal Data Servers (PDS)}\label{sec:pds}

A PDS stores repositories and associated media files, and allows anybody to query the data it hosts via a HTTP API.
Moreover, a PDS provides a real-time stream of updates for the repositories it hosts via a WebSocket.
Indexers (see Section~\ref{sec:indexing}) subscribe to this stream in order to find out about new or deleted records (posts, likes, follows, etc.) with low latency.
This architecture is illustrated in Figure~\ref{fig:indexing}.

Hosting a PDS for a small number of users requires only small computing resources, even if those users have a large number of followers.
Users who wish to self-host their own PDS can therefore do so on a cheap virtual machine in the cloud, or even on a Raspberry Pi connected to their home internet router.
However, we expect that most users will sign up for an account on a shared PDS run by a professional hosting provider~-- either Bluesky Social PBC, or another company.

Compared to choosing a Mastodon server, the user's choice of PDS hosting provider is fairly inconsequential.
The PDS URL is internal to the system, and is not normally visible to users.
It makes no difference whether two users are on the same PDS or different PDSes, since interaction between users goes via the indexing infrastructure in any case.
A user can migrate from one PDS to another by copying their repository and media files to the new PDS, and pointing their account at the new PDS URL (Section~\ref{sec:identity}).
Even if a PDS shuts down without warning, users can upload a backup of their repository to a new PDS, and thus recover their account without losing any of their posts or their social graph.

PDS operators generally perform some basic moderation by deleting any illegal content hosted on their servers.
However, PDS-level moderation is much less important than server-level moderation in Mastodon, because in atproto, the primary moderation role is taken on by separate actors in the system~-- the labelers and feed generators (see Section~\ref{sec:labeling}).
This allows different sets of people to offer server hosting and moderation services, respectively; we believe this separation is valuable since operating a server and moderating a community require largely disjoint sets of skills~\cite{Roth:2023}.

\subsection{Indexing Infrastructure}\label{sec:indexing}

On the web, websites are crawled and indexed by search engines, which then provide web-wide search and discovery features that the websites alone cannot provide.
The AT Protocol is inspired by this architecture: the repositories hosted by PDSes are analogous to websites, and the indexing infrastructure is analogous to a search engine.
User repositories are primary data (the ``source of truth''), and the indexes are derived from the content of the repositories.

At the time of writing, most of Bluesky's indexing infrastructure is operated by Bluesky Social PBC (indicated by a shaded area in Figure~\ref{fig:indexing}).
However, the company does not have any privileged access: since repositories are public, anybody can crawl and index them using the same protocols as our systems use.
If Bluesky Social PBC were to violate users' expectations~-- for example, by censoring (omitting from the index) accounts that users wish to see~-- other parties would be free to provide their own indexes that do not perform this censorship.
Client apps could switch to reading from a different index, or even use a combination of multiple indexes.

In July 2024 (when Bluesky had 6 million users), it was possible to maintain a real-time copy of all user repositories on a single server for US\$153 per month~\cite{NewboldRelay}~-- expensive for a hobby project, but easily affordable by many organizations.
This figure includes repository storage and the inbound bandwidth for fetching records from PDSes, but not the computational resources to build and serve summary indexes from that data.

Due to the higher cost, we expect that there will be fewer hobbyist indexers than self-hosted PDSes.
Nevertheless, as Bluesky grows, there are likely to be multiple professionally-run indexers for various purposes.
For example, a company that performs sentiment analysis on social media activity about brands could easily create a whole-network index that provides insights to their clients.
Web search engines can incorporate Bluesky activity into their indexes, and archivists such as the Internet Archive can preserve the activity for posterity.

The indexing infrastructure operated by Bluesky Social PBC is illustrated in Figure~\ref{fig:indexing}.
It is composed of multiple services that have integration points for external services.

\subsubsection{The Relay}\label{sec:relay}

The first component is the \emph{Relay}, which crawls the user repositories on all known PDSes and consumes the streams of updates that they produce.
The Relay checks the signatures and Merkle tree proofs on updates, and maintains its own replica of each repository.
From this information, the Relay creates the \emph{firehose}: an aggregated stream of updates that notifies subscribers whenever records are added or deleted in any of the known repositories.
Normally the WebSocket connection to PDSes provides the relay with low-latency notifications of repository changes, but in case of network interruptions the relay can also periodically re-crawl repositories and compare them to its local replica to determine what records have been added or deleted.

The firehose is publicly available.
Consuming the firehose is an easier way of building an index over the whole network, compared to directly subscribing to the source PDSes, since the Relay performs some initial data cleaning (discarding malformed updates, filtering out illegal content and high-volume spam).
The firehose includes Merkle proofs and signatures along with records, allowing subscribers to check that they are authentic.

The Relay does not interpret or index the records in repositories, but simply stores and forwards them.
Any developers wanting to create a new social mode on top of atproto can define a new lexicon with new record types, and these records can be stored in existing repositories and aggregated in the firehose without requiring any changes to the Relay.

\begin{figure*}
    \centering
    \includegraphics[width=\linewidth]{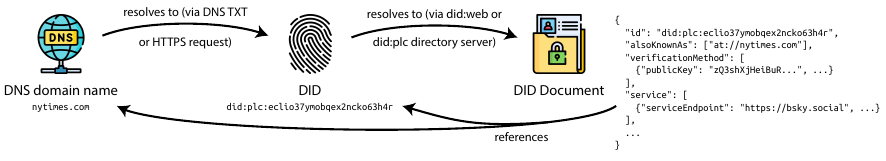}
    \Description{Handle (e.g. nytimes.com) resolves to DID via DNS TXT record or HTTPS request. DID resolves to DID document via did:web or did:plc directory server.}
    \caption{A handle resolves to a DID, and a DID resolves to a DID document, which in turn references the handle, DID, and the user's public key. Icons from Flaticon.com.}
    \label{fig:identity}
\end{figure*}

% Icons used:
% https://www.flaticon.com/free-icon/dns_2333476
% https://www.flaticon.com/free-icon/fingerprint-scan_6692271
% https://www.flaticon.com/free-icon/folder_4883508

\subsubsection{The App View}\label{sec:appview}

The App View is a service that consumes the firehose, and processes the records that are relevant to the Bluesky social app (records in the \texttt{com.atproto} and \texttt{app.bsky} lexicons).
For example, the App View counts the number of likes on every post, and it collates the thread of replies to each post.
The App View also maintains the set of followers for each user, and constructs the timeline containing the posts by the accounts that each user is following.
It then offers a web service through which this information can be queried.
When a record contains references to images, the App View fetches those files from the original PDS, resizes them if necessary to reduce the file size, and makes them available via a content delivery network (CDN).

To display this information in the user's client app, the client queries the user's own PDS, which then fetches the necessary data from an App View.
% Devin confirms this is correct. The reason it goes through the PDS is to authenticate the request to the App View, so the App View can verify which user it's serving.
The App View is also responsible for enforcing moderation controls: for example, if one user has blocked another, and one of the users' repositories contains a record of an interaction that should not have been allowed due to the block, then the App View drops that interaction so that nobody can see it in the client apps.
This behavior is consistent with how blocking works on Twitter/X~\cite{TwitterBlock}, and it is also the reason why blocks are public records in Bluesky: every protocol-conforming App View needs to know who is blocking who in order to enforce the block~\cite{PublicBlocks,PrivateBlocks}.
If users are unhappy with the moderation rules applied by the App View operated by Bluesky Social PBC, it is always possible for third parties to operate alternative App Views that index the same firehose and present the data in a different way.

Other social modes built upon atproto require their own App View services to interpret and index the records in users' repositories as required.
The PDSes and relay can be reused.

% Push notifications need to go through a service run by the developer of the client app.
% For the official Bluesky app, the app view handles push notifications.
% For 3rd-party clients we have an open interface for any client to register a user with their own push notification service, but at the moment they must generate the notifications themselves, e.g. by following the firehose. 

\subsection{Labelers and Feed Generators}\label{sec:labeling}

Relay and App View aim to provide a mostly ``unopinionated'' service: they compute indexes over repositories in a neutral way, without attempting to rank or classify content.
However, a good user experience also requires ``opinionated'' judgements for the purposes of content filtering (e.g.\ detecting sexually explicit images or spam) and curation (e.g.\ selecting posts on a particular topic).

The AT Protocol separates out the ``opinionated'' aspects of the system into separate services: \emph{labelers} and \emph{feed generators}.
These services typically take the firehose as their input.
Labelers produce a stream of judgements about content (e.g.\ ``this post is spam''), whereas feed generators return a list of post IDs they have selected for inclusion in a custom feed.
Users can choose in their client app which feeds and which labelers they want to use.
The output of labelers is consumed by App Views or PDSes in order to apply content filtering~\cite{Labeling}.
For a feed generator, an App View expands the post IDs into full posts before sending them to the client app of users who have subscribed to that feed.

Having labeler and feed generator services that are separate from App Views has several advantages:
\begin{itemize}
    \item Anyone can run such services, which enables a pluralistic ecosystem in which different parties may make different judgements about the same piece of content.
        Users, as well as the operators of App Views and PDSes, can decide whose judgements they want to trust, and it is easy for them to switch to alternative labeling and feed generation services if their current providers fail to meet their expectations.
    \item It becomes easier to set up alternative App View providers: since any App View can consume the publicly available output from labelers and feed generators, there is less pressure for each App View to develop its own content filtering infrastructure.
        Having alternative App Views is important for a healthy, decentralized marketplace.
\end{itemize}

Feed generators can be implemented using our starter kit~\cite{FeedGeneratorKit}, or using a third-party service such as Skyfeed~\cite{Skyfeed}.

\subsection{User Identity}\label{sec:identity}

The DNS-based user identity in atproto is decoupled from hosting and indexing.
Anybody owning a domain name can issue subdomains as atproto handles, and this does not require running a PDS or any other service besides DNS.

Handles can change, but every Bluesky/atproto account also has an immutable, unique identifier: a \emph{decentralized ID} or \emph{DID}, which is a URI starting with the prefix ``\texttt{did:}''.
For example, when a record in user $A$'s repository indicates that $A$ is following $B$, that record contains $B$'s DID; this allows a user to change their handle without affecting their social graph.
DIDs are a recent W3C standard~\cite{DIDCore}.

Moreover, we want a user to be able to migrate to a different PDS without changing either their DID or their handle.
DIDs provide a mechanism for \emph{resolving} a DID into a \emph{DID document}, a JSON document containing information about the user identified by that DID, as illustrated in Figure~\ref{fig:identity}.
In atproto, a DID document specifies the handle of the user, the URL of their PDS, and the public key that is used to sign the Merkle tree root of their repository every time they add or delete a record.
To change their handle or their PDS, the user needs to update their DID document to the new value.

To prove ownership of a handle, the user must have a bidirectional link between their DID and their domain name handle, as shown in Figure~\ref{fig:identity}:
\begin{itemize}
    \item A link from the handle to the DID is established either by storing the DID in a DNS TXT record on that domain name, or by returning the DID in response to a HTTPS request to a \texttt{/.well-known/} URL on that domain name~\cite{DomainHandle}.
    \item A link from the DID to the handle is established by including the handle in the DID document that is returned when the DID is resolved.
\end{itemize}
The App View periodically checks these bidirectional links, and invalidates the handle if either is broken.
Provided that the App View is honest and takes measures to protect against DNS poisoning attacks (perhaps using DNSSEC when available), this approach prevents users from impersonating domains that they do not own.

\subsubsection{Resolving DID documents}

The DID specification~\cite{DIDCore} does not directly specify the mechanism for resolving a DID into a DID document.
Rather, the first substring after \texttt{did:} in a DID indicates the \emph{DID method}, and the specification of the DID method defines the protocol for obtaining the DID document.
Hundreds of DID methods have been defined~\cite{DIDMethods}, many of which are dependent on specific blockchains or other external systems.
To avoid atproto implementations having to support so many resolution methods, our services currently only accept DIDs based on either \texttt{did:web} (defined by the the W3C Credentials Community Group~\cite{did:web}) or \texttt{did:plc} (defined by ourselves for atproto~\cite{did:plc}).
Support for more DID methods might be added in the future.

The \texttt{did:web} method is very simple: the part of the DID after \texttt{did:web:} is a domain name, and the DID document is resolved by making a HTTPS request to a \texttt{/.well-known/} URL on that domain name.
The security of a \texttt{did:web} identity therefore assumes that the web hosting provider for that domain is trusted, and also relies on trusting the TLS certificate authorities that may authenticate the HTTPS request.

\texttt{did:web} identities are therefore similar to domain name handles, with the difference that the name cannot be changed, since a DID is an immutable identifier.
This makes \texttt{did:web} appropriate for the identity of organizations that are already strongly linked to a particular domain name.
For most users, \texttt{did:plc} is more appropriate, since it uses a domain name only as a handle that can be changed.

\subsubsection{The did:plc DID method}

When a user creates an account on the Bluesky social app, they are by default assigned a DID of the form \texttt{did:plc:eclio37ymobqex2ncko63h4r}, where the string after the prefix \texttt{did:plc:} is the SHA256 hash of the initial DID document, truncated to 120 bits and encoded using base32~\cite{did:plc}.
A DID of this form can be resolved to the corresponding DID document by querying a server at \url{https://plc.directory/}, which is currently operated by Bluesky Social PBC; in the future we plan to establish a consortium of independent operators that collectively provide the PLC directory service.

The PLC directory server plays an authoritative role similar to the DNS root servers, but it is mostly untrusted because PLC DID documents are self-certifying.
If the DID document has not changed since its initial creation, it is easy to verify that a DID has been correctly resolved to a DID document by recomputing its hash.
To support changes to the DID document, the initial version of a user's DID document contains a public key that is authorized to sign a new version of the DID document.
Any new version of the DID document is only valid if it has been signed by the key in the previous version.
The directory returns all DID document versions for a given DID, allowing anybody to check the chain of signatures.

If the directory server were to be malicious, it would not be able to modify any DID documents~-- it could only omit valid DID document versions from its responses, or fail to respond at all.
Moreover, if there were to be a fork in DID document history such that two correctly signed successor versions for some DID document exist, the directory server could choose which one of these forks to serve.
To mitigate such attacks, we anticipate that a future version of the PLC directory will use an append-only transparency log similar to certificate transparency~\cite{Laurie:2014}.

\subsubsection{Authentication}

In principle, the cryptographic keys for signing repository updates and DID document updates can be held directly on the user's devices, e.g.\ using a cryptocurrency wallet, in order to minimize trust in servers.
However, we believe that such manual key management is not appropriate for most users, since there is a significant risk of the keys being compromised or lost.

The Bluesky PDSes therefore hold these signing keys custodially on behalf of users, and users log in to their home PDS via username and password.
This provides a familiar user experience to users, and enables standard features such as password reset by email.
The AT Protocol does not make any assumptions about how PDSes authenticate their users; other PDS operators are free to use different methods, including user-managed keys.

\ifextendedversion
\section{Related Work}\label{sec:related-work}

Several other decentralized social networks are in development.
We believe that there is no single optimal design: different systems make different trade-offs, and are therefore suitable for different purposes.
Bluesky and the AT Protocol aim to provide a good user experience by providing a global view over the whole network, making moderation a first-class concern, and having clients that are lightweight and easy to use.
For example, conversation threads include all replies (unless removed by moderation), regardless of the server on which they were posted.
To achieve this goal we rely on an indexing infrastructure that is more centralized than some other designs.
However, we emphasize that there can be multiple competing indexers, and third-party client apps are free to show data from whichever indexers they wish.

In 2021 some of our team published a review of the decentralized social ecosystem~\cite{EcosystemReview}.
In this section we summarize some recent developments that have happened since, and we refer to the review for a more comprehensive comparison of protocols and systems.

Many decentralized social networking projects have ideas in common.
For example, the idea of using DNS domain names as usernames also appears in Nostr~\cite{NostrDNS}.
An atproto PDS has similarities to Git repository hosting (e.g.\ GitHub/Gitlab) or a Solid pod~\cite{Solid}.
% There are also federated chat systems such as Matrix~\cite{Matrix}, IRC~\cite{IRC}, and XMPP~\cite{XMPP}, but we focus on systems that provide a Twitter-like model where users follow each other.

\subsection{Scuttlebutt}

Secure Scuttlebutt (SSB) is a peer-to-peer social networking protocol~\cite{Scuttlebutt}; Manyverse~\cite{Manyverse} is a social application built upon the SSB protocol.
It optionally uses relay servers called \emph{pubs} to store messages from peers that are offline, and to enable user discovery.
The client software downloads the feeds from accounts that the user is explicitly following, and from accounts followed by followed accounts (up to three hops by default).
This can require significant amounts of storage and bandwidth on the client.

Any messages from users outside of the third-degree network are not shown, which effectively limits the set of people who can mention or reply to a user to the third-degree network.
This deliberate design decision is intended to reduce moderation problems by prioritizing conversation between people who already know each other~\cite{ManyverseBluesky}.
In contrast, Bluesky/atproto are designed to allow anybody to talk to anybody else.
This requires more explicit moderation to manage unwanted content, but we believe it also enables serendipity and is a prerequisite for any ``digital town square''.

Since SSB is built upon append-only logs and gossip replication, it is not possible to delete content once it has been posted~\cite{SSBDeletion}.
User identity is tied to a cryptographic key on the user's device, requiring manual key management for moving to another device.
Posting from multiple devices is not possible, as sharing the same key between devices can make an account unrecoverable~\cite{SSBMultiDevice}.
A successor protocol to SSB, called PZP, is designed to address these issues~\cite{PZP}.

\subsection{Nostr}

Nostr also began as a revision of SSB, replacing the append-only logs with individual signed messages~\cite{SSBNostr}.
It leans more heavily on relay servers instead of peer-to-peer communication: clients publish and fetch messages on relays of their choice, and there is no federation among relays~\cite{Nostr}.
The protocol is deliberately simple, and it prioritizes censorship resistance over moderation: relays can block users, but users can always move to a new relay, and use multiple relays at the same time.
Communication (e.g.\ reply threads) is only possible between users who have at least one relay in common.
Although some services index the whole Nostr network, these indexes are not used for user-to-user interaction.
As a result, it is unpredictable who will see which message.
The creator of Nostr writes: ``there are no guarantees of anything [\dots] to use Nostr one must embrace the inherent chaos''~\cite{NostrVision}. 
Key management is manual in Nostr, and facilities for key rotation are still under discussion~\cite{NostrKeyRotation}.

% Are like counts and follower counts approximate? NIP-45 defines server-side aggregation
% https://github.com/nostr-protocol/nips/issues/159
% https://github.com/nostr-protocol/nips/blob/master/45.md

\subsection{Farcaster and blockchain-based systems}

Farcaster~\cite{Farcaster} has some architectural similarities to Bluesky/atproto, although it was developed independently.
It has storage servers called \emph{hubs}, which store the state of the entire network similarly to an atproto Relay, and it has a concept of \emph{hosted app servers} that are similar to our App View~\cite{FarcasterOverview}.
Farcaster user IDs are similar to our DIDs, and they are mapped to public keys using a smart contract on the Ethereum Optimism blockchain that is functionally similar to our PLC directory.
Usernames can be either ENS names~\cite{ENS}, or names within an off-chain namespace managed centrally by Farcaster, similarly to \texttt{.bsky.social} subdomains in Bluesky~\cite{FarcasterArchitecture}.

A difference is that Farcaster has no equivalent to atproto's PDS; instead, client apps publish signed messages directly to a hub, and hubs synchronize messages using a convergent gossip protocol.
Users must pay in cryptocurrency to register their public key, and for hub data storage (at the time of writing, Ethereum equivalent to \$5~USD/year); when a user exceeds their storage allowance, old messages are deleted.
Fees are currently collected centrally by the Farcaster team~\cite{FarcasterFees}.
In contrast, the AT Protocol does not specify storage limitations, but leaves it to providers of PDS and indexing services to define their own business model and abuse-prevention policies.
We also prefer to avoid a dependency on a cryptocurrency.

The Lens protocol~\cite{Lens} is more strongly blockchain-based than Farcaster: it even stores high-volume user actions such as posts and follows on Polygon, a proof-of-stake blockchain.
DSNP takes a similar approach~\cite{DSNP}.
Placing high-volume events directly on a blockchain incurs orders of magnitude higher per-user costs than atproto, and is likely to run into scalability limits as the number of users grows.
Lens is adopting a layer-3 blockchain that provides better scalability and lower cost~\cite{LensMomoka}, but weaker security properties.
Linking social accounts to cryptocurrency wallets and NFTs enables users to monetize their content, but this is not a goal of atproto.

\subsection{ActivityPub and Mastodon}

ActivityPub~\cite{ActivityPub} is a W3C standard for social networking, and Mastodon~\cite{Mastodon} is its most popular implementation.
Mastodon gives a lot of power to server administrators: for example, a server admin can choose to block another server, preventing all communication between users on those servers. %, even those who were not involved in the disagreement that led to the blocking.
There is a degree of lock-in to a server because moving to another server is intrusive: the username changes, moving posts to the new server currently requires an experimental command-line tool~\cite{MastodonPostMigration,MastodonContentMover}, and other users' replies to those posts are lost.
If the old server is not reachable~-- for example, because its admin shut it down without warning or because its domain was seized~\cite{QueerAF}~-- the user's social graph is lost.
These risks can be mitigated by self-hosting; managed providers exist~\cite{MastodonHosting}, but they still require some expertise and cost money.
The AT Protocol separates the roles of moderation and hosting, and aims to make it easier to change providers without losing any data.

When user $A$ follows user $B$, $A$'s server asks $B$'s server to send it notifications of $B$'s future posts via ActivityPub.
This architecture has the advantage of not requiring a whole-network index.
However, replies to a post notify the server of the original poster, but not necessarily every server that has a copy of the original post, leading to inconsistent reply threads on different servers.
Notifications can be forwarded, but in the limit this leads to each server having a copy of the whole network, which would make it expensive to run a server.
Viral posts can generate a lot of inbound requests to a server from people liking, replying, and boosting (reposting).
In comparison, the Bluesky indexing infrastructure is also fairly expensive, but a PDS is cheap to run.
Since users can choose their moderation preferences independently from their indexing provider (App View), we believe that the ecosystem can be healthy with a small number of indexing providers.

% Centralising tendency: for new users who don't have a dedicated server for their specific community, the default choice is to join one of the big servers.
\fi % \ifextendedversion

\section{Conclusions}

Bluesky and the AT Protocol are a new approach to social media.
Their architecture is based on the principle that every part of the system can be run by multiple competing providers, and users can switch providers with minimal friction (in particular, without changing username, and without losing any of their content or social graph).
For example, anyone can write a client, host a PDS, index the network, or provide moderation services, and all of these services interoperate.
Even though the majority of Bluesky services are currently operated by a single company, we nevertheless consider the system to be decentralized because it provides \emph{credible exit}~\cite{NewboldProgress}: if Bluesky Social PBC goes out of business or loses users' trust, other providers can step in to provide an equivalent service using the same dataset and the same protocols.

% The AT Protocol is also extensible to other social modes besides microblogging.

While some decentralized systems prioritize censorship resistance, we believe that a good user experience requires explicitly addressing problematic content such as harassment and misinformation.
We therefore make moderation a first-class concern that is handled separately from infrastructure hosting, and we provide strong mechanisms for users to control the content they see.
Our open architecture allows a pluralistic system in which different users may choose different providers that uphold different values, while still allowing them to communicate and interoperate.

% there is no global consensus on what content is acceptable.
% A user on a self-hosted or loosely moderated PDS may post controversial content, but they are not entitled to the attention of others: App Views may choose not to index the content, and clients may ignore it depending on the user's moderation settings.
% This philosophy is sometimes described as ``free speech, but not free reach''~\cite{DiResta:2018}.

% The AT Protocol provides cryptographically authenticated data, but our implementation pairs it with custodial key management to provide a familiar user experience.

% 3rd-party clients allow bsky pbc to offer a more curated, opinionated service with stronger moderation, protecting bsky brand
% e.g. bsky own labelers mandatory
% clients may bundle moderation defaults, e.g. enforcing usage of a particular labeler
% can't ban 3rd-party clients like reddit apollo
% prior art: content that is illegal e.g. in germany: twitter filters it out client-side

% Do we want to talk about some kind of robots.txt-like mechanism through which users can specify preferences regarding crawling and indexing?

% Do we want to include some sort of discussion about business models? How do the operators of app views, labelers, etc. get paid?

% \begin{acks}
    % Martin Kleppmann is supported by the Volkswagen Foundation and crowdfunding supporters including Mintter and SoftwareMill.
% \end{acks}

\bibliographystyle{ACM-Reference-Format}
\balance
\bibliography{references}
\end{document}